\documentstyle[]{mn}
\newif\ifAMStwofonts

\input{epsf}

\newcommand{\Halpha}{H$\alpha$}
\newcommand{\Halpham}{{\rm H}\alpha}

\newcommand{\degrees}{$^{\circ}$}

\newcommand{\ergcms}{erg\,cm$^{-2}$s$^{-1}$}
\newcommand{\kms}{~km\,s$^{-1}$}

\newcommand{\IRAF}{{\scriptsize IRAF}}

\newcommand{\DAOPHOT}{{\scriptsize DAOPHOT}}

\newcommand{\sigmar}{\sigma_{R}}
\newcommand{\sigmaphi}{\sigma_{\phi}}
\newcommand{\sigmaz}{\sigma_{z}}
\newcommand{\sigmalos}{\sigma_{los}}
\newcommand{\hs}{h_{\sigma}}
\newcommand{\othree}{[O~{\sc iii}]}
\newcommand{\hone}{H~{\sc i}}
\newcommand{\htwo}{H~{\sc ii}}

\ifoldfss
  \ifCUPmtlplainloaded \else
    \NewTextAlphabet{textbfit} {cmbxti10} {}
    \NewTextAlphabet{textbfss} {cmssbx10} {}
    \NewMathAlphabet{mathbfit} {cmbxti10} {} 
    \NewMathAlphabet{mathbfss} {cmssbx10} {} 
  \fi
  \ifAMStwofonts
    \ifCUPmtlplainloaded \else
      \NewSymbolFont{upmath} {eurm10}
      \NewSymbolFont{AMSa} {msam10}
      \NewMathSymbol{\upi}     {0}{upmath}{19}
      \NewMathSymbol{\umu}     {0}{upmath}{16}
      \NewMathSymbol{\upartial}{0}{upmath}{40}
      \NewMathSymbol{\leqslant}{3}{AMSa}{36}
      \NewMathSymbol{\geqslant}{3}{AMSa}{3E}

    \fi
  \fi
\fi 

\ifnfssone
  \newmathalphabet{\mathit}
  \addtoversion{normal}{\mathit}{cmr}{m}{it}
  \addtoversion{bold}{\mathit}{cmr}{bx}{it}
  \newmathalphabet{\mathbfit} 
  \addtoversion{normal}{\mathbfit}{cmr}{bx}{it}
  \addtoversion{bold}{\mathbfit}{cmr}{bx}{it}
  \newmathalphabet{\mathbfss} 
  \addtoversion{normal}{\mathbfss}{cmss}{bx}{n}
  \addtoversion{bold}{\mathbfss}{cmss}{bx}{n}
  \ifAMStwofonts
    \ifCUPmtlplainloaded \else
      \UseAMStwoboldmath
      \makeatletter
      \new@mathgroup\upmath@group
      \define@mathgroup\mv@normal\upmath@group{eur}{m}{n}
      \define@mathgroup\mv@bold\upmath@group{eur}{b}{n}
      \edef\UPM{\hexnumber\upmath@group}
      \new@mathgroup\amsa@group
      \define@mathgroup\mv@normal\amsa@group{msa}{m}{n}
      \define@mathgroup\mv@bold\amsa@group{msa}{m}{n}
      \edef\AMSa{\hexnumber\amsa@group}
      \makeatother
      \mathchardef\upi="0\UPM19
      \mathchardef\umu="0\UPM16
      \mathchardef\upartial="0\UPM40
      \mathchardef\leqslant="3\AMSa36
      \mathchardef\geqslant="3\AMSa3E
    \fi
  \fi
\fi 

\ifnfsstwo
  \DeclareMathAlphabet{\mathbfit}{OT1}{cmr}{bx}{it}
  \SetMathAlphabet\mathbfit{bold}{OT1}{cmr}{bx}{it}
  \DeclareMathAlphabet{\mathbfss}{OT1}{cmss}{bx}{n}
  \SetMathAlphabet\mathbfss{bold}{OT1}{cmss}{bx}{n}
  \ifAMStwofonts
    \ifCUPmtlplainloaded \else
      \DeclareSymbolFont{UPM}{U}{eur}{m}{n}
      \SetSymbolFont{UPM}{bold}{U}{eur}{b}{n}
      \DeclareSymbolFont{AMSa}{U}{msa}{m}{n}
      \DeclareMathSymbol{\upi}{0}{UPM}{"19}
      \DeclareMathSymbol{\umu}{0}{UPM}{"16}
      \DeclareMathSymbol{\upartial}{0}{UPM}{"40}
      \DeclareMathSymbol{\leqslant}{3}{AMSa}{"36}
      \DeclareMathSymbol{\geqslant}{3}{AMSa}{"3E}
    \fi
  \fi
\fi 

\ifCUPmtlplainloaded \else
  \ifAMStwofonts \else 
    \def\upi{\pi}
    \def\umu{\mu}
    \def\upartial{\partial}
  \fi
\fi

\title[PNe Kinematics in M94]{Using Slitless Spectroscopy to study the
Kinematics of the Planetary Nebula Population in M94
}

\author[N.G. Douglas, J. Gerssen, K. Kuijken and M.R. Merrifield]
	{N.~G.~Douglas,$^1$
	J. Gerssen,$^1$
	K. Kuijken,$^1$
	M.R. Merrifield,$^2$\\
	$^1$Kapteyn Astronomical Institute, Groningen, Netherlands\\
	$^2$School of Physics \& Astronomy, University of Nottingham, U.K.}

\pagerange{\pageref{firstpage}--\pageref{lastpage}}
\pubyear{1999}

\begin{document}

\maketitle

\label{firstpage}

\begin{abstract}
 The planetary nebula populations of relatively nearby galaxies can be
easily observed and provide both a distance estimate and a tool with
which dynamical information can be obtained.  Usually the requisite
radial velocities are obtained by multi-object spectroscopy once the
planetary nebulae have been located by direct imaging.  Here we report
on a technique for measuring planetary nebula kinematics using the
double-beam ISIS spectrograph at the William Herschel Telescope in a
novel slitless mode, which enables the detection and radial velocity
measurements to be combined into a single step.  The results on our
first target, the Sab galaxy NGC 4736, allow the velocity dispersion of
the stellar population in a disk galaxy to be traced out to four scale
lengths for the first time and are consistent with a simple isothermal
sheet model.  
 \end{abstract}


\section{Introduction}

The outer kinematics of galaxies have played a crucial role in our
understanding of their structure. The dark matter halos are most
important there, so that conclusions about their shape, mass and
extent may be drawn that are less dependent on assumed mass-to-light
ratios of the observed stars. Most of the angular momentum resides at
large radii, and relaxation times are longest there, possibly enabling
echos of the formation process to be observed directly.  However, the
required observations are rather difficult.  The integrated stellar
light of a galaxy rapidly becomes too weak at large radii to do
spectroscopy.  In the case of elliptical galaxies, the old stellar
populations have now in a few cases been probed as far as two effective
radii (e.g., Carollo et al. 1995; Gerhard et al. 1997). 

Some tracers, such as globular clusters and \hone\ emission, can be
observed at larger radii, but neither provides a reliable tracer of the
kinematics of the relaxed, old stellar population.  Moreover systems
like S0s and ellipticals generally lack an extensive gaseous disk.
Fortunately an alternative tracer of the kinematics out to large radii
has been identified by Hui et al.~(1993), who showed that the radial
velocities of a galaxy's planetary nebula (PN) population constitute a suitable
diagnostic. Planetary nebulae (PNe) appear in the post-main sequence phase of
stars in the range 0.8 - 8 M$_{\odot}$. Fortunately, 
in all but the very youngest of systems 
the PN population is strongly correlated with the older, and therefore
dynamically relaxed, population of low-mass stars. This statement
is true not only
because of the statistics of stellar formation and evolution, but
also because 
the PN lifetime is itself a strongly decreasing function of progenitor mass
(Vassiliadis \& Wood 1994).

PNe emit almost all of their light in a few bright emission lines,
particularly the \othree\ line at 5007\AA. There is evidence that the PN
\othree\ luminosity function is essentially constant with galaxy type
and metallicity (Jacoby et al 1992), so that the observed bright-end
cutoff magnitude $M^*$ (Ciardullo et al 1989) represents a `standard candle' with
which distances can be determined.  At a distance of 10~Mpc, this
cutoff corresponds to a flux of $2.0 \times 10^{-16}$
\ergcms, making PNe
within one dex of this limit easily detectable in one night with a 4-m
telescope. As a rule of thumb, approximately 100 such PNe are found to
be present per $10^9$ L$_{\odot}$ of B-band luminosity (Hui
1993), so they are seen in sufficient number to study the kinematics
of the stellar population of the host galaxy.

The usual approach in making such kinematic studies has been to identify
the PN population by narrow-band imaging and then to re-observe the
detected PNe spectroscopically to obtain radial velocities.  However,
other strategies exist that avoid the need for several observing runs. 
For example, Tremblay et al.~(1995) used Fabry-Perot measurements in a
PN kinematics study of the SB0 galaxy NGC~3384.  In this paper, we
describe a novel alternative, based on slitless spectroscopy, and
discuss its application to the Sab galaxy M94.

\section{Detection and Kinematics of PN{\sc e} through Slitless Spectroscopy}

Our method for obtaining the kinematics of PNe is outlined in
Figure~\ref{fig:schematic}. 
The galaxy under study is imaged through narrow-band filters around 
the two strongest emission lines in a typical PN spectrum, \Halpha\ and
\othree.  The \Halpha\ image is recorded directly, but the \othree\
light is dispersed. Comparison of the dispersed and undispersed images
then allows the kinematics of the PNe to be measured, without prior
knowledge of the location of the PNe. Two modes of analysis are possible:
\begin{figure}
\epsfxsize\hsize
\epsfbox{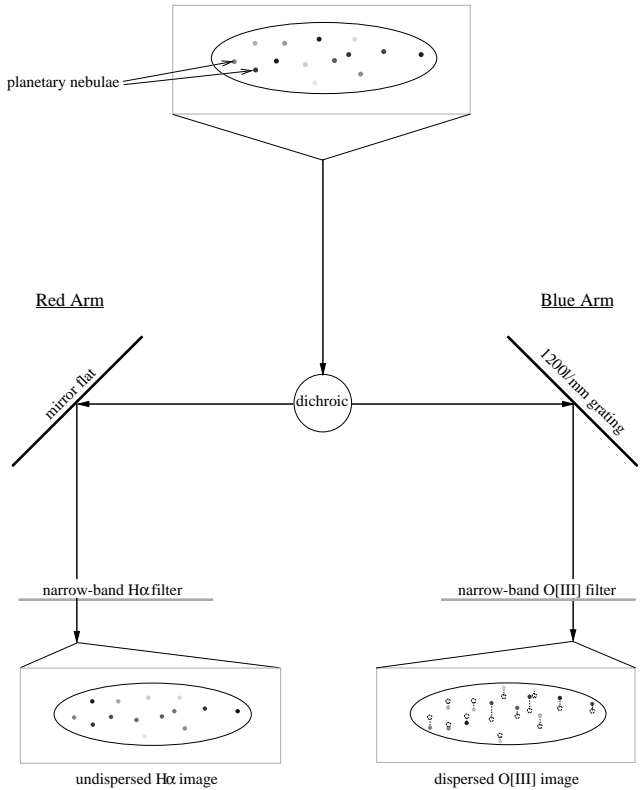}
\caption{Diagram showing schematically how one can use a
dual-beam spectrograph to study PN kinematics.  The narrow-band images in
the two arms allow one to identify the PNe and measure their
line intensities in both the \othree\ 5007\AA\ and \Halpha\ lines.  The
dispersive element in the blue arm shifts each PN image by an amount
proportional to its redshift. In the red arm the grating has been replaced by a
mirror.}
\label{fig:schematic}
\end{figure} \\
{\bf Dispersed/Undispersed Imaging (DUI):} In the dispersed blue arm the
PNe will be visible through their \othree\ emission as point sources,
displaced from their `true' positions by an amount related to their
radial velocity, against a background of dispersed galactic light.  The
red arm will detect the PNe through their \Halpha\ emission, along with
any other objects with line or continuum emission in the pass band of
the filter.  Assuming that the PNe can be unambiguously identified in
the \Halpha\ image, their position in the \othree\ image will give
the radial velocity. 
 We chose to disperse the blue rather than the red
light since PN have a higher flux at \othree\ than at \Halpha, and
gratings are less efficient than mirrors. \\
{\bf Counter-Dispersed Imaging (CDI):} The method of
counter-dispersed imaging was described in an earlier paper (Douglas
\& Taylor 1999). In this mode, pairs of dispersed \othree\ images are
made with the entire spectrograph rotated by 180 degrees between
exposures. The difference in the position of a given PN in the two
dispersed images again reflects its radial velocity.  In this case the
undispersed \Halpha\ image can be used as a consistency check on the
derived positions of the PNe.

We have implemented our method on the ISIS 
medium-dispersion spectrograph at the Cassegrain (f/10.94) focus of the 4.2m
William Herschel Telescope.  The slit unit was removed during the
observations, and the \othree\ and \Halpha\ light paths were separated
with a dichroic before being passed through appropriate narrow-band
filters. The filters ($\lambda$5026/47 and $\lambda$6581/50) were
custom-made for this project in order to exploit the full $4 \times 1$
arcmin field of the instrument in slitless mode, and to give adequate
velocity coverage.  We used a 1200 g/mm (first order) grating in the
blue arm. Both arms contained
1024$^2$-pixel  Tek CCD detectors.  Thus, we obtained dispersed
images in the blue (calibration showed the dispersion to be about
24\kms\ per pixel) and simultaneous direct images in the red. DUI
observations are accomplished in a single exposure; CDI requires two
exposures.

The overall efficiency of this setup,
including telescope, instrument, filter and CCD,
was found from observations of a
standard star (Feige 34) to be 14\% in the blue and 20\% in the red
for air mass of 0.
We therefore expected to detect $\sim 2.7$ \othree\ photons per second
from the brightest PN when viewed at 6 Mpc.  Dark sky (V=21.4) would
produce $\sim$2 counts per arcsec$^2$ per second, and the background
light of the galaxy 1--5 counts per arcsec$^2$.  A reasonable goal is
to obtain a 4$\sigma$ detection  of
the PN population 
over the top decade (2.5 mag) of
the luminosity function. This requires  $\sim$ 4 hours of integration if
the seeing conditions are of the order of one arcsec.  The required
integration time is approximately proportional to the square of
the seeing.

\section{Observations}

The observations were carried out on 1997 April 11 and 12.  For this
pilot project NGC~4736 (M94) was chosen. It has a large angular size
and is at a distance of 6 Mpc (Bosma et al 1977).  The galaxy shows
some peculiar morphological features, most notably an inner and an
outer optical ring with radii of 1 and 5.5 arcmin respectively.
The stellar light is too faint for direct optical spectroscopy outside
 1 arcmin, and our goal was to measure PN kinematics to
three times this radius. Key parameters of NGC~4736 are listed in
Table~\ref{gpars}, and the observing log is given in Table~\ref{ISISlog}.

\begin{table}
\caption{Galaxy parameters}
\scriptsize
\begin{center}
\begin{tabular}{lcc}
\hline
Name & NGC 4736 (M94) \\
\hline
Position (J2000) & 12h50m53.061s \\
                 & +41d07m13.65s\\
Hubble type & SA(r)ab \\
V$_{\rm Helio}$ & 308 $\pm$ 1 km/s \\
B$_{\rm T}^0$ & 8.99 \\
Angular size & $11.2 \times 9.1$~arcmin \\
Inclination & $\approx$ 35 $^\circ$  \\
Scale length & 57 $\pm$ 10~arcsec& \\
Position angle  & 105$^\circ$ \\
\hline
\end{tabular}
\end{center}
\label{gpars}
\end{table}

Two fields were observed, 3 arcmin west of centre on the major axis,
and 3 arcmin north of centre on the minor axis. The major axis was
observed at two orientations (allowing CDI mode) while the minor axis
was observed in one orientation  only. We took the major axis position angle
to be 90\degrees, as would seem appropriate from  the relevant isophotes
(see Figure~\ref{Field}).
Total integration times were 6.0hrs on
the Western field (4.6hrs in one orientation, and 1.4hrs with the
spectrograph rotated by 180 degrees), and 3hrs on the Northern field.
The observing conditions were close to photometic with seeing, as
judged from stellar images in the red arm, less than 1.1\arcsec\ at
all times.  We also observed a flux standard star for photometry and a
Galactic PN as a radial velocity reference.  A second star was
observed as a spectral reference.

The custom-made \othree\ and \Halpha\ filters had a central wavelength and
peak transmission of 5026\AA/0.823 and 6581\AA/0.915, respectively.
The nominal FWHM was 47\AA\ and 50\AA, while
the effective photometic bandwidth 
was evaluated graphically and found to be 38.7\AA\  
and 45.7\AA, respectively.

\begin{table}
\caption{Log of our observations at the WHT, La Palma Observatory.
PA is
 the position angle of the spectrograph and T the  total integration time 
 (where two values are given these refer to the red/blue arms
respectively). 
}
\scriptsize
\begin{center}   
\begin{tabular}{|l|r|r|r|r|}
\hline
Field         &PA               & T & UT & Airmass\\
\hline
	& & &  April 11 & \\
HD66637                & 90  &60        & 20:35 & 1.06 \\         
Feige 34               & 90  &2/10      & 21:15  & 1.07\\
N4736 - major axis      & 90 &16,453        &  21:20 -  2:13 & 1.35 - 1.07\\
PN 49+88.1              & 90 &    5         & 2:03  & 1.06\\
N4736  - major axis     & 270&  4,868   &   2:36 -  4:52 & 1.12 - 1.54\\
&&&&\\
        & & &  April 12 & \\
HD66637                & 90  &0.1/10 & 20:29 & 1.06\\
Feige 34               & 90  &3/60       &  20:57  & 1.09\\
N4736 - minor axis     & 0   &10,800       &  21:02 - 0:08 & 1.42 - 1.03\\
\hline
\end{tabular}
\end{center}

\label{ISISlog}
\end{table}

\section{Data Reduction}
\subsection{Calibration}
\label{calibration}

The dispersion in the blue arm was measured by inserting a slit and
using an arc lamp, and found to be 0.3992\AA/pixel.  The spectrum of
the star HD66637 was then observed through the same slit and
wavelength-calibrated.  Subsequent observations of the same star at
numerous positions in the field (after removal of the slit)
established that the dispersion could be taken as constant over the
field.  The combination of these observations with the undispersed red
arm positions of HD66637 gave an unambiguous solution for the
transformation between objects in the red (direct image) and blue
(dispersed image).  (Note that with this technique the
radial velocity of the star does {\em not} enter into the calculation.)

To check the zero point of the velocity scale, we moved the telescope
from the reference star to the Galactic planetary nebula PN~49.3+88.1,
for which the heliocentric radial velocity is listed as $-141$\kms\ 
(Schneider et. al 1983).  In ten pointings over the field of the spectrograph we
measured $-136.0\pm 3.2$\kms, in agreement with the calculated
observatory frame redshift of $-133$\kms.  Unfortunately we discovered
later that at certain telescope orientations the flexure is large
enough to cause significantly larger errors.  
However, such flexure only introduces an offset in absolute velocity, and  
does not compromise our ability to study a galaxy's internal kinematics.
In order to derive an absolute calibration for the velocity scale, we 
later obtained a long slit observation of two of the objects detected in
this analysis (see \S~\ref{longslit}).

The spectrograph
field of view with the slit unit removed consists of an approximately
unvignetted area of about 4\arcmin $\times$ 1\arcmin, but we obtained
useful data outside this region. Correcting the observed fluxes for the
vignetting is only straightforward for the red (undispersed) arm. In
the blue arm the correction is complicated by the fact that the
image is dispersed. The sky flat measured in the blue arm was found to
closely approximate the aperture
function in the red arm, shifted by a small number of pixels,
transformed to blue arm coordinates and then convolved with the filter
profile at the appropriate dispersion.

We therefore carried out the complete analysis after correcting only for
the pixel-to-pixel variation of the CCD responses, determined in the
usual way.  Once the PNe were identified it was possible to determine
their wavelength and their positions in the aperture prior to being
shifted by the spectrograph, so that the \othree\ magnitudes could then
be corrected analytically both for the filter response, which was fitted
with a polynomial, and for vignetting.

\subsection{Object identification}
\label{oi}

Scripts based on \IRAF\ procedures were used for all of the data reduction.  

Due to spectrograph flexure, individual exposures needed to be aligned
before being added. This was accomplished by a simple shift (at least
one PN was visible in each individual exposure).  The images were then
combined by computing a weighted and scaled median.  A spatial median filter
was applied to the combined frame, and the result subtracted from the
original image to yield a field of unresolved objects against a
background with a mean of zero.  

\begin{figure}
\epsfxsize=\hsize
\epsfbox{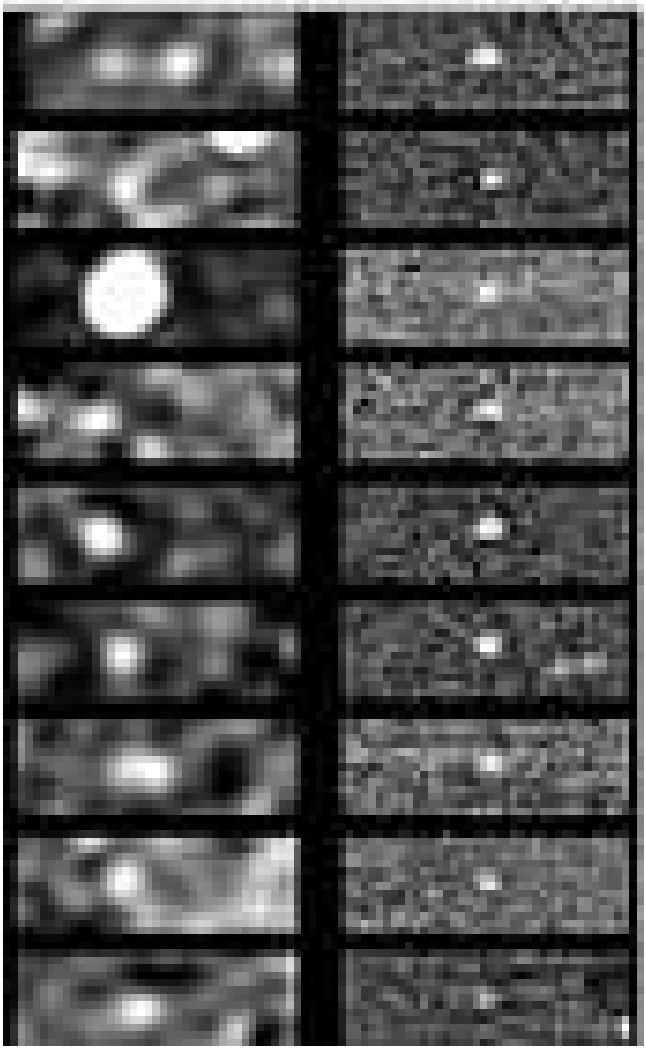}
\caption{An example of the `postage stamp' method for detecting red
counterparts. The boxes on the right each contain one point
source identified in the blue frame. The boxes on the left side show
the corresponding  parts of the red frame. Any red counterpart must be located
somewhere within the box since the size of the box is such that it
encompasses the velocity range spanned by the filter. The horizontal
location of the counterpart is a measure of the object's radial
velocity. In the vertical direction we require the positional
agreement to be
within one pixel.}

\label{stamps}
\end{figure}

The PNe and any other point sources in the red and blue images were
extracted by two methods:\\ 
(A) Blinking and hand-tagging, followed
by a PSF-fitting step to evaluate the shape and size parameters.
For DUI mode observing it was usually found to be easier to search
the range of possible (red) coordinates corresponding to an object
seen in the blue (see Fig.~\ref{stamps}).  \\
(B) an automated procedure based on object lists generated with
\DAOPHOT.  A 2D-gaussian fit to each detected image was used to select
PN candidates.  The FWHM of a candidate was required to be within a small range
of that of the seeing disk (PNe are unresolved) and the axial ratio
close to the value 1.28 expected from our instrumental configuration
(the ellipticity arises from the anamorphic effect of the
grating). The object lists were then correlated to search for potential PN
image pairs.

\subsection{False detections}
\label{fd}

As well as PNe, our observations will also detect other objects having a
predominantly line-emission spectrum near the \othree\ line.  In the case
of a bright spiral galaxy like M94 it is obvious that \htwo\ regions will
mimick PNe.  As they belong to a younger population than the PNe their
inclusion in the analysis would lead to an underestimate of the velocity
dispersion.  Although it is tempting to use the \othree /\Halpha\ line
ratio as a discriminant (it tends to be larger in PNe) this is not sharp
enough to eliminate the \htwo\ regions without eliminating a fair fraction
of the PNe.  Unless a better discriminant can be found, this source of
contamination may ultimately restrict the viability of our technique to the
elliptical and S0 galaxies for which it was devised.  However in the
particular case of M94 the problem is manageable because:

(i) M94 is sufficiently close that a large fraction of bright \htwo\
regions would be spatially extended and thus eliminated by one of the
object selection criteria; (ii) the \htwo\ regions in M94 are mostly
confined to features associated with spiral arms - one of these passes
through the eastern extremity of our major axis field and all objects
there were ignored.  The rest of the field, and the minor axis field,
are devoid of recognised \htwo\ agglomerations; (iii) the total number of
unresolved objects detected in \othree\ is close to the number expected
for PNe on the basis of their luminosity specific density as seen in
other galaxies. 
 
None of these factors eliminates the problem entirely, but their
combined effect is such that we feel that contamination is negligible. 
Even if 10\% of the PNe had been misidentified the effect on the
calculated velocity dispersion would be at most 5\%, well
below other sources of error. 

High redshift galaxies, in which the Ly$\alpha$ line is shifted into
the \othree\ passband, form another potential source of contamination
when long integrations are made of extended halos (Freeman et al.\ 1999).  This
is also not a significant issue in the present case.

\subsection{Comparing Spectral Modes}

We identified PNe in NGC~4736 along the minor axis with DUI mode
observations and along the major axis using CDI (with DUI mode data
being redundant). Although CDI requires two distinct integrations, we
found that data obtained from CDI results in a higher number of
detected PN per unit integration time. For visual identification using
blinking, CDI is considerably easier since the two images have the
same plate scale and similar properties with respect to sky noise and
confusion. Therefore to illustrate the numerical superiority of using
two dispersed \othree\ images we rely on the automated search results
for the major axis observations only. 36 PNe were identified in this
field from matching 2.5$\sigma$ sources in the counterdispersed
\othree\ images. The limiting factor here was the shorter integration
time with one of the two spectrograph orientations: had both
integration times been equal we would presumably have found yet more
PNe. By comparison, only 24 PNe were detected from the DUI mode
analysis of the same field. Therefore, we conclude that a significant
number of PNe are too faint in \Halpha\ for the DUI mode to detect
them.

\section{Long-slit data}
\label{longslit}

 It has been mentioned that instrument flexure, particularly between
sets of observations in CDI mode, can lead to an uncertainty in the
absolute velocity scale.  To remove this uncertainty we attempted to
obtain the velocity of at least one object in the major axis field via
the Willian Herschel Telescope service data program.  As the PNe are
faint, with effective V-band magnitude of around 25, the slit had to be
positioned `blind' on the basis of the position computed from the
dispersed data.  Fortunately the astrometry does not depend on velocity,
but only on correct identification of PN pairs, and on the centroiding
of the dispersed images of stars in the field. 

 We requested a spectrum using ISIS in long-slit mode and with the slit
at a position and PA chosen such as to fall across {\em two} objects. 
This provided a good test of the astrometry. 
 The service observation was attempted on 1999 July 28.  Both objects
were acquired, and their separation along the slit agreed with that
calculated.  One of the objects, suspected of being an \htwo\ region, was
confirmed as such.  The radial velocities obtained had an internal error
of about 10\kms, as judged from the values obtained from different
lines, and were used to calibrate the major axis data (Table
\ref{PNpa90}).

\section{Results}
\label{results}

\begin{figure}

\epsfxsize=100mm
\epsfbox{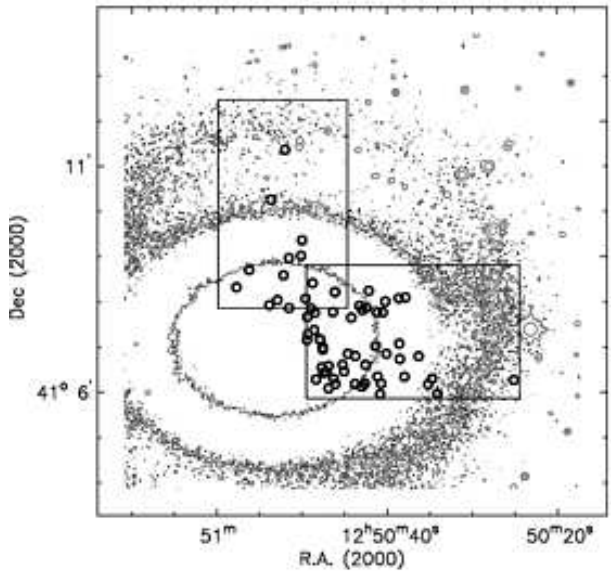}
\caption{Digitised Sky Survey image of NGC~4736 (North at top)
with the observed fields marked with boxes and the PNe
shown as open circles.
}
\label{Field}
\end{figure}

The PNe identified along the major axis are listed in Table~\ref{PNpa90}
and those along the minor axis in Table~\ref{PNpa0}.  Their positions
are also shown in Fig~\ref{Field}.  As suggested by the successful
long-slit experiment, the positional uncertainty is of the order of 1-2
arcsec.  The internal error in the velocities is approximately 10\kms.
 The minor axis velocities have an offset that has
not been determined, but the values as presented have an average
velocity near systemic, as would be expected.  The derivation of
velocities from DUI mode data, as was used for the minor axis, is in
fact much less sensitive to flexure.  However, in order to reduce the
number of candidate objects in the red image, only radial velocities
between 200 and 500\kms\ were searched for, so this table has to be used
with caution.

\begin{table}
\caption{Fifty-three point sources in the major axis field.
 Rvel  is the relative (heliocentric)
radial velocity measured from the slitless spectroscopy plus an
offset established by an additional calibration (see \S\ref{longslit}). 
$m_{5007}$ = -2.5 log($F$) - 13.74 is the apparent V-band magnitude
of the objects, where $F$ is the flux in \ergcms. The magnitudes have been
corrected for vignetting (see \S\ref{calibration}).  The vignetting
correction could not be reliably applied for two objects as they were
too close to the edge of the aperture, and their magnitudes are not
given. 
FWHM (in pixels) refers to the major axis of
the 2D-gaussian fit.  } 
\scriptsize
\begin{center}
\begin{tabular}{lllll}
\hline
\hline
RA (J2000)   & Dec (J2000)       & Rvel      & $m_{5007}$ & FWHM \\
\hline
12:50:49.46 & 41:07:09.5 & 459  & 23.44  & 3.52    \\   
12:50:49.42 & 41:07:40.3 & 472  & 23.18  & 3.59    \\   
12:50:49.35 & 41:07:17.5 & 463  & 23.48  & 3.31    \\   
12:50:48.93 & 41:07:51.2 & 483  & 22.54  & 3.29    \\   
12:50:48.62 & 41:07:22.9 & 444  & 23.18  & 3.31    \\   
12:50:48.40 & 41:06:17.3 & 309  & 24.80  & 3.24    \\   
12:50:48.53 & 41:07:45.8 & 450  & 24.63  & 3.54    \\   
12:50:47.97 & 41:07:10.8 & 508  & 25.46  & 3.9    \\   
12:50:47.87 & 41:07:09.8 & 378  & 25.78  & 2.79    \\   
12:50:47.73 & 41:06:33.2 & 296  & 24.56  & 3.9    \\   
12:50:47.61 & 41:07:00.2 & 345  & 24.45  & 3.61    \\   
12:50:47.59 & 41:06:56.9 & 391  & 25.33  & 3.38    \\   
12:50:47.50 & 41:06:26.2 & 525  & 25.52  & 3.5    \\   
12:50:47.17 & 41:06:27.4 & 429  & 25.34  & 3.54    \\   
12:50:46.92 & 41:06:05.5 & 413  & 25.14  & 3.45    \\   
12:50:46.93 & 41:06:35.3 & 378  & 25.41  & 3.47    \\   
12:50:46.32 & 41:06:18.3 & 357  & 25.60  & 3.05   \\   
12:50:46.40 & 41:07:46.8 & 466  & 24.95  & 4.38    \\   
12:50:46.13 & 41:06:11.2 & 380  & 24.24  & 3.94    \\   
12:50:46.17 & 41:08:13.6 & 491  & 24.34  & 3.57    \\   
12:50:45.27 & 41:06:36.8 & 478  & 24.91  & 4.37    \\   
12:50:45.11 & 41:06:27.8 & 308  & 25.30  & 3.73    \\   
12:50:44.65 & 41:06:51.3 & 369  & 24.82  & 3.66    \\   
12:50:44.26 & 41:07:39.2 & 432  & 25.78  & 2.67    \\   
12:50:43.79 & 41:06:11.4 & 353  & 24.95  & 3.33    \\   
12:50:43.76 & 41:06:48.3 & 348  & 25.18  & 2.79    \\   
12:50:43.29 & 41:07:54.9 & 457  & 22.59  & 3.52    \\   
12:50:43.03 & 41:06:07.7 & 485  & 24.91  & 3.17    \\   
12:50:42.95 & 41:07:49.5 & 473  & 26.39  & 2.7    \\   
12:50:42.66 & 41:06:11.8 & 405  & 22.65  & 3.92  \\   
12:50:42.58 & 41:06:36.6 & 428  & 25.00  & 4.34    \\   
12:50:42.48 & 41:07:52.8 & 425  & 24.88  & 3.5    \\   
12:50:42.19 & 41:08:15.0 & 476  & 24.58  & 3.07    \\   
12:50:41.40 & 41:07:01.5 & 433  & 24.98  & 3.5    \\   
12:50:41.17 & 41:06:21.0 & 413  & 25.21  & 3.47    \\   
12:50:41.17 & 41:07:46.4 & 339  & 24.26  & 4.7    \\   
12:50:40.87 & 41:05:57.5 & 414  &        & 3.47    \\   
12:50:40.74 & 41:06:11.7 & 410  & 25.53  & 4.37    \\   
12:50:40.55 & 41:07:46.4 & 417  & 24.72  & 3.61   \\   
12:50:40.26 & 41:08:00.7 & 400  & 25.33  & 4.04    \\   
12:50:40.15 & 41:06:51.3 & 383  & 24.74  & 2.77    \\   
12:50:38.62 & 41:07:04.9 & 409  & 25.41  & 3.71    \\   
12:50:38.67 & 41:08:05.0 & 391  & 25.35  & 4.2    \\   
12:50:38.55 & 41:06:44.4 & 418  & 25.00  & 2.89    \\   
12:50:38.02 & 41:06:20.9 & 423  & 25.54  & 3.68    \\   
12:50:37.90 & 41:08:06.1 & 464  & 25.14  & 3.05    \\   
12:50:36.34 & 41:06:48.0 & 430  & 25.65  & 3.07    \\   
12:50:35.24 & 41:06:10.8 & 425  & 24.93  & 3.45    \\   
12:50:34.75 & 41:06:18.3 & 408  & 24.87  & 3.31    \\   
12:50:34.15 & 41:05:58.2 & 379  &        & 2.91    \\   
12:50:25.18 & 41:06:16.4 & 415  & 25.50  & 4.3    \\   
12:50:48.71 & 41:06:32.7 & 378  & 25.18  & 5.45    \\   
12:50:42.79 & 41:07:48.9 & 454  & 25.22  & 3.54    \\   
\hline
\hline
\end{tabular}
\end{center}
\label{PNpa90}
\end{table}

\begin{table}
\caption{Fourteen point sources in the minor axis field. The
coordinates are derived directly from the position in the red image
(see text). 
$\rm\frac{\Halpham}{\othree}$ gives the ratio of the fluxes
(in \ergcms) seen in the red and blue images.  Other symbols
are as in Table~\ref{PNpa90}, except that the radial velocities
have not been checked against an external reference and may 
have a small common offset. }
\scriptsize
\begin{center}
\begin{tabular}{lllccl}
\hline
\hline
RA (J2000)   & Dec (J2000)       & Rvel   
                    & $m_{5007}$ & $\rm\frac{\Halpham}{\othree}$ & FWHM \\
\hline
12:50:52.02 & 41:11:22.3 & 330  & 26.26  & 1.26  & 3.36   \\
12:50:53.68 & 41:10:15.8 & 263  & 24.37  & 0.38  & 3.17   \\
12:50:50.03 & 41:09:22.2 & 341  & 26.56  & 1.11  & 3.71   \\
12:50:50.16 & 41:09:01.5 & 311  & 25.67  & 0.82  & 4.04   \\
12:50:51.57 & 41:08:58.3 & 270  & 24.74  & 0.62  & 3.1   \\
12:50:56.25 & 41:08:42.6 & 265  & 25.54  & 0.65  & 4.74   \\
12:50:52.22 & 41:08:35.6 & 268  & 24.35  & 0.48  & 3.54   \\
12:50:48.81 & 41:08:25.3 & 367  & 26.21  & 1.57  & 2.72   \\
12:50:57.74 & 41:08:19.7 & 252  & 25.35  & 0.65  & 4.25   \\
12:50:46.14 & 41:08:13.2 & 408  &    	&   	& 3.61  \\
12:50:49.62 & 41:08:04.5 & 363  & 24.84  & 1.28  & 5.02  \\
12:50:52.93 & 41:08:02.5 & 496  & 25.63  & 2.47  & 4.46   \\
12:50:53.86 & 41:07:56.1 & 310  & 23.35  & 2.61  & 3.31   \\
12:50:51.57 & 41:07:52.1 & 367  & 23.62  & 8.87  & 3.76   \\
 \hline
\hline
\end{tabular}
\end{center}
\label{PNpa0}
\end{table}

\subsection{Luminosity function}
\label{pnlf}
\begin{figure}
\epsfxsize=\hsize
\epsfbox{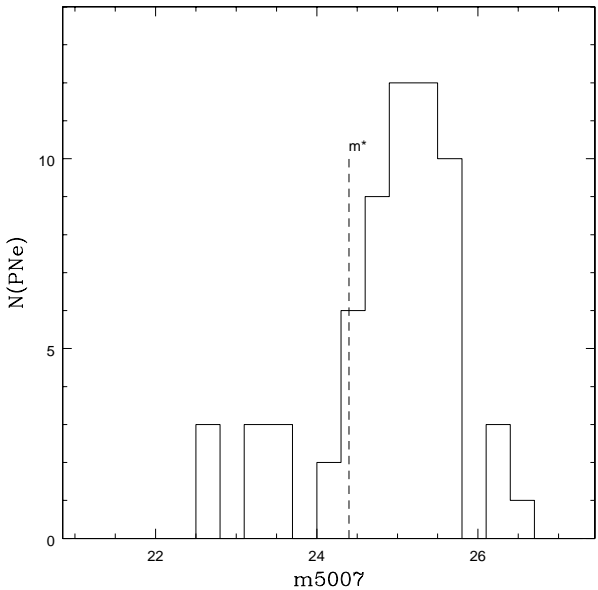}
\caption{The luminosity function of the 64 objects for which magnitudes
could be determined. The cutoff peak for PNe is 
clearly defined. The objects brighter than this cutoff are probably bright
\htwo\ regions}
\label{majlf}
\end{figure}

We placed a premium on detecting as many PNe as possible, even
in the partially vignetted region of the instrument. Considerable
corrections have been applied,
and the magnitudes should therefore only be taken as indicative.
The luminosity function of the objects detected is presented in
Fig~\ref{majlf}. The bright-end cutoff for the assumed distance of 6~Mpc
($m^{*}$ = 24.4)
is indicated.  At the faint end the luminosity function is, of
course, significantly incomplete while at the bright end some objects
are brighter than the cutoff.  The latter are probably \htwo\ regions and
have therefore not been included in the analysis of the kinematics.

The number of PNe found in the major axis field (CDI mode) is in rough
agreement with predictions.  From the basic data on NGC~4736 compiled by
Mulder (1995) ($m_B = 8.58$, $D = 6.0$~Mpc) we have $L_B =
2.07 \times 10^{10} L_{\odot}$, and on the basis of the results of
Hui et al (1993) the expected number of PN in the top decade of the PNLF in
M94 would therefore be around 2000.  Mulder also found 
the galaxy to be fairly well-fitted by
an exponential disk with scale length $h = $57\arcsec.  The region we
examined includes 0.029 of the light of such an exponential, which
should therefore include 59 PNe in the brightest decade. This number compares
well with the 53 actually found, though the agreement may be somewhat
fortuitous given the incompleteness at the faint end.

\subsection{Rotation curve}
\label{rc}

\begin{figure}

\epsfxsize=\hsize
\epsfbox{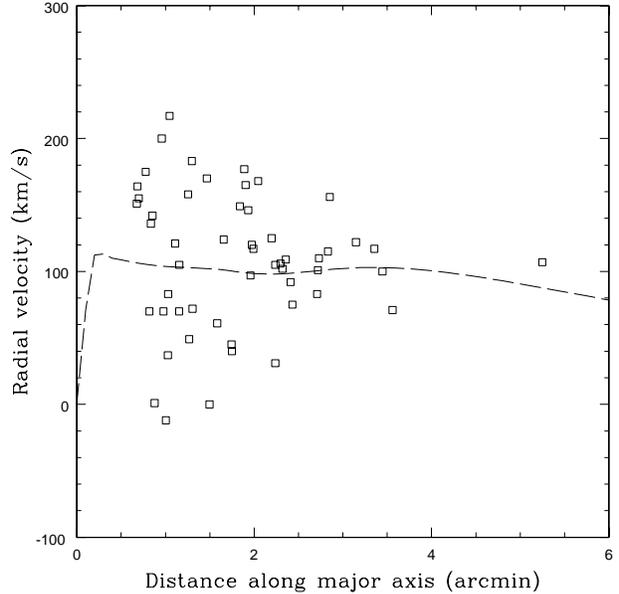}
\caption{Line-of-sight velocities of the 53 objects in the major axis 
field plotted as a function 
of distance along the axis. 
The dotted line shows the gas rotation curve, from Sofue (1997) }
\label{majvrad}

\end{figure}

In Fig~\ref{majvrad} the line-of-sight velocities of the 53 objects in
the major axis field are plotted as function of radius, after
subtraction of the systemic velocity (Table~\ref{gpars}).  Flat rotation
is seen out to the last measured point at almost 5 scale lengths.  For
comparison we overplot the \hone/CO rotation curve of Sofue (1997),
projected into the plane of the sky.  For objects near a distance of 1
arcmin along the major axis the mean velocity is 98\kms, in agreement
with the projected gas rotation velocity of 103\kms\ at that point.  The
uncorrected minor axis data yield a mean velocity of 329\kms, consistent
with the systemic velocity.

\subsection{Velocity dispersion}

Generally, the vertical structure in disks of spiral galaxies is
reasonably well described by an isothermal sheet approximation (van der
Kruit \& Searle 1982; Bottema 1993).  In this model, the vertical
velocity dispersion is found to follow an exponential decline with
radius with scale length twice that of the surface density.  With the
additional assumption that the dispersion ellipsoid has constant axis
ratios throughout the disk, one finds that the line-of-sight velocity
dispersion follows the same decline, independent of the galaxy's
inclination.  We tested this using the major axis data. 
Figure~\ref{majdisp} shows the velocity dispersion in bins of distance
along the major axis.  Seven objects were eliminated from the kinematic
analysis as their brightnesses suggested that they are \htwo\ regions. 
The curve is the least-squares exponential fit, yielding central
velocity dispersion 111 \kms\ and scale length $\hs$= 130 arcsec.  These
are close to the published value for the central stellar velocity
dispersion ($120 \pm 15$ \kms) obtained from absorption-line spectra
(Mulder and van Driel, 1993) and to twice the photometric scale length
($2h = 114$ arcsec), suggesting that the isothermal sheet approximation
is reasonable. 

\subsection{Combined Kinematic Model}

\begin{figure}

\epsfxsize=\hsize
\epsfbox{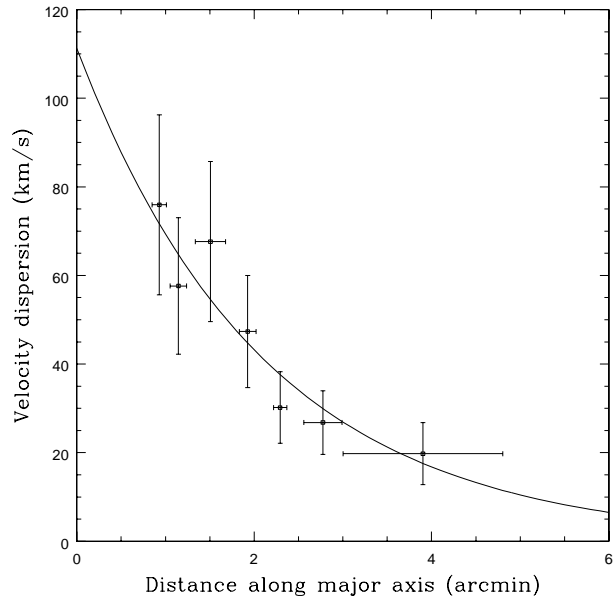}
\caption{The dispersion in radial velocities computed by binning  
the radial velocity measurements, together with the best-fit
exponential curve}

\label{majdisp}
\end{figure}

The binning in Fig~\ref{majdisp} effectively assumes that the PNe all
lie close to the major axis.  In fact, they are located up to one
arcminute from the axis, at azimuths up to 45\degrees, so a more
sophisticated approach is required. We therefore projected the PNe on to
a thin disk of fixed inclination (35\degrees), giving
$r,\phi$ coordinates.  The nebulae's line-of-sight velocities can then
be compared with a model consisting of a three-dimensional isothermal
sheet with a flat rotation curve.  This model has five parameters,
namely the three components of the central velocity dispersion
$\sigmaz,\sigmaphi,\sigmar$, the scale length $\hs$, and the rotation
amplitude. A maximum likelihood method
was then used to fit the model.  We added twelve PNe from the
minor axis field (Table~\ref{PNpa0}) to help constrain the fit (two
were excluded from the fit as probable \htwo\ regions).

In practice the data were not adequate to constrain all five
parameters.  Using the canonical relationship
$\sigmaphi^2=\sigmar^2/2$ from the epicyclic approximation (Binney \&
Merrifield 1998, eq.~11.18) and allowing $\sigmaz/\sigmar$ to vary
over the range 0.2 to 2.0, we found a robust
 maximum likelihood solution with 
scale length $\hs = 144 \pm 30$~arcsec, 
central
velocity dispersion $\sigmalos = 120 \pm 30 $\kms, and
circular rotation speed $v_c = 177 \pm 11$\kms. These results are consistent
with the  \hone\ rotation speed at 1~arcmin 
radius (180 \kms) and again with twice the 
photometric scale length. 

It was not
possible to constrain the shape of the
velocity ellipsoid with these data -- such an analysis 
would require a more complete azimuthal coverage of the galaxy.
However if we assume $0.5 < \sigmaz/\sigmar < 1.0$
then we infer $75 < \sigmar < 110$\kms\ at one  photometric scale length,
consistent with the trend between rotation speed and
disk velocity dispersion found by Bottema (1993).

Thus far we have ignored measurement error in the velocities, which will
tend to increase the measured velocity dispersion. This turns out to
be a small effect: allowing for a 1$\sigma$ error of 10\kms, the
fitted dispersion becomes approximately 3\% smaller and the scale
length is unchanged.

\section{Conclusions}

In this paper we have demonstrated how the kinematics of the PN
population in a galaxy can be measured by slitless spectroscopy
through narrow-band filters with a dual-beam spectrograph. We compared
two possible modes: dispersed/undispersed imaging, in which a
dispersed \othree\ image is compared to an undispersed \Halpha\ image;
and counterdispersed imaging, in which two \othree\ images, dispersed
in opposite directions, are analysed. It turns out that the latter
method is more effective: evidently the \Halpha\ fluxes of faint PNe are
not reliably high enough to allow both spectral lines to be used.

Our pilot experiment was performed on the large nearby Sab galaxy
M94. It has revealed a PN population in the disk whose rotation curve
remains flat, and whose velocity dispersion declines radially
exponentially, consistent with the predictions of a simple isothermal
sheet model. PNe were detected out to five exponential scale lengths,
well beyond the reach of kinematic measurements based on
integrated-light absorption-line spectroscopy. The number of PNe
detected was consistent with expectations.

The present experiment was limited to two fields in this large
galaxy. Complete coverage of the galaxy should yield around 2000 PNe,
and would allow a detailed kinematic model to be fitted, including a
determination of the axis ratio of the velocity ellipsoid following
the technique of Gerssen et al. (1997).  Obtaining such data for a
small sample of nearby galaxies in just a few nights of 4-m telescope
time is a practical proposition.

\section{Acknowledgements}

The WHT is operated on the island of La Palma by the Isaac Newton Group
in the Spanish Observatorio del Roque de los Muchachos of the Instituto
de Astrof\'isica de Canarias.  We wish to acknowledge the help and
support of the ING staff.  We are also grateful for some excellent
additional data provided by ING astronomers in service mode.  The {\sf
IRAF} data reduction package is written and supported by the IRAF
programming group at the National Optical Astronomy Observatories (NOAO)
in Tucson, Arizona.  We thank the referee Dr R.  Ciardullo for comments
which led to the addition of \S\ref{fd}.

\label{lastpage}

\end{document}